\documentclass[aps,prd,preprint,floatfix]{revtex4}
\usepackage{epsfig}

\begin{document}

\preprint{\rightline{ANL-HEP-PR-02-016}}

\title{Lattice QCD at finite isospin density at zero and finite temperature.}

\author{J.~B.~Kogut}
\address{Dept. of Physics, University of Illinois, 1110 West Green Street,
Urbana, IL 61801-3080, USA}
\author{D.~K.~Sinclair}
\address{HEP Division, Argonne National Laboratory, 9700 South Cass Avenue,
Argonne, IL 60439, USA}

\begin{abstract}
We simulate lattice QCD with dynamical $u$ and $d$ quarks at finite chemical
potential, $\mu_I$, for the third component of isospin ($I_3$), at both zero
and at finite temperature. At zero temperature there is some $\mu_I$, $\mu_c$
say, above which $I_3$ and parity are spontaneously broken by a charged pion 
condensate. This is in qualitative agreement with the prediction of effective
(chiral) Lagrangians which also predict $\mu_c=m_\pi$. This transition
appears to be second order, with scaling properties consistent with the
mean-field predictions of such effective Lagrangian models. We have also
studied the restoration of $I_3$ symmetry at high temperature for $\mu_I >
\mu_c$. For $\mu_I$ sufficiently large, this finite temperature phase
transition appears to be first order. As $\mu_I$ is decreased it becomes
second order connecting continuously with the zero temperature transition.
\end{abstract}
\maketitle

\pagestyle{plain}
\parskip 5pt
\parindent 0.5in

\section{Introduction}

Neutron stars are made of dense cold nuclear matter -- hadronic matter at high
baryon-number density and low temperature. Large nuclei can be considered as
droplets of nuclear matter. The relativistic heavy-ion collisions now being
observed at RHIC and the CERN heavy-ion program can produce hadronic matter at
high temperature and finite baryon-number density. Nuclear matter also has a
finite (negative) isospin ($I_3$) density due to Coulomb interactions, and it
has been suggested that at very high densities it would have a finite
strangeness density. Hence it is of interest to study QCD at finite
quark/baryon-number density, finite isospin density and finite strangeness
density at both zero and finite temperature.

Finite density is customarily studied through the introduction of a chemical
potential for the charge of interest in the action. Introducing a finite
chemical potential for quark-number leads to a complex fermion determinant 
with a real part of indefinite sign, which precludes use of standard simulation 
methods which rely on importance sampling. If we include a finite chemical
potential, $\mu_I$, for $I_3$ in the absence of any quark-number chemical
potential, the fermion determinant remains non-negative, and simulations are
possible. Such simulations can determine the QCD phase structure on one 
surface in the phase diagram for nuclear matter. One can hope that this will
identify phases which will persist to finite baryon/quark-number density, and
determine their properties.

We have performed simulations of lattice QCD with 2 flavours of light staggered
quarks at finite $\mu_I$ for both zero and finite temperatures. Preliminary
results of these simulations were reported at Lattice2001 \cite{lattice2001}.
We included a small explicit $I_3$-breaking interaction needed to observe
spontaneous symmetry breaking on a finite lattice. In addition to allowing us
to observe spontaneous breaking of $I_3$ (and parity), this term renders the
fermion determinant strictly positive. Our zero temperature simulations were
performed on an $8^4$ lattice at an intermediate value of the coupling
constant. In the limit that our symmetry-breaking parameter vanishes, there is
some $\mu_I=\mu_c$ above which $I_3$ and parity are broken spontaneously by a
charged pion condensate. This is in accord with the predictions of Son and
Stephanov using effective (chiral) Lagrangians \cite{sonstep}. We observe
critical scaling consistent with the mean-field predictions of these effective
(chiral) Lagrangians. In drawing these conclusions, it is important to use
equations of state of the forms predicted by such effective Lagrangians.
These results are similar to those obtained in studies of the quenched version
of this theory \cite{quenched}. They are also similar to what is observed for
2-colour QCD at finite chemical potential $\mu$ for quark-number
\cite{qcd2us,qcd2it,qcd2jp,newqcd2}. This is not surprising since the
effective Lagrangian analysis of 2-colour QCD at finite $\mu_I$ \cite{cpt2} is
similar to that for QCD at finite $\mu_I$.

Our finite temperature simulations were performed on $8^3 \times 4$ lattices.
At sufficiently high temperature and $\mu_I > \mu_c$, we observe the 
evaporation of the symmetry-breaking pion condensate. For $\mu_I$ sufficiently
large, this transition is first order. As $\mu_I \rightarrow \mu_c$ this
transition softens and appears to become second order. Such a transition from
first to second order should occur at a tricritical point. Again these results
are similar to what we observed for 2-colour QCD at finite quark-number 
chemical potential \cite{qcd2t}.

Section 2 gives details of the actions and their symmetries. In section 3 we
present our zero temperature results and scaling analyses. The finite 
temperature results are presented in section 4. Discussions, conclusions and
an outline of planned extensions are given in section 5.

\section{Lattice action and symmetries}

The staggered fermion part of the action for lattice QCD with degenerate $u$
and $d$ quarks at a finite chemical potential $\mu_I$ for isospin ($I_3$) is
\begin{equation}
S_f=\sum_{sites} \bar{\chi}[D\!\!\!\!/(\tau_3\mu_I)+m]\chi
\end{equation}
where $D\!\!\!\!/(\mu)$ is the standard staggered $D\!\!\!\!/$ with links in
the $+t$ direction multiplied by $e^{\frac{1}{2}\mu}$ and those in the $-t$
direction multiplied by $e^{-\frac{1}{2}\mu}$ \cite{quenched}.
When $\mu_I=m=0$, this action has a global $U(2) \times U(2)$ flavour symmetry
under which
\begin{eqnarray}
\chi & \longrightarrow & \exp[i(\theta+\epsilon\phi).\tau]\chi \nonumber \\
\bar{\chi} & \longrightarrow & \bar{\chi}\exp[-i(\theta-\epsilon\phi).\tau]
\end{eqnarray}
where $\tau=(1,\vec{\tau})$, $\theta$ and $\phi$ are site-independent
4-component ``vectors'', and $\epsilon = \epsilon(x) = (-1)^{x+y+z+t}$.
Spontaneous symmetry breaking can occur in any direction in this space. If we
keep $\mu_I=0$ and allow $m \ne 0$, the symmetry is broken down to $U(2)_V$.
On the other hand if we keep $m=0$ and allow $\mu_I \ne 0$, the symmetry is
broken down to $U(1) \times U(1) \times U(1) \times U(1)$ generated by $1$,
$\tau_3$, $\epsilon$ and $\epsilon\tau_3$. Finally in the general case where
neither $\mu_I$ nor $m$ vanishes, the symmetry is reduced to $U(1)_V \times
U(1)_V$ associated with $1$ and $\tau_3$.

In order to predict potential symmetry breaking patterns we make several simple
modifications of the arguments of Son and Stephanov \cite{sonstep} to apply
them to the staggered lattice action. The generic quark bilinear which creates
a meson has the form
\begin{equation}
M = \bar{\chi}\Gamma\chi.
\end{equation}
The propagator for such a meson obeys the inequality
\begin{equation}
|\langle M(x) M^\dagger(0) \rangle| \le 
                     const \langle S(x,0) S^{\dagger}(x,0)\rangle.
\end{equation}
Thus, meson operators $M$ whose propagators are proportional to 
$\langle S(x,0) S^{\dagger}(x,0)\rangle$, are potential Goldstone bosons.
Now we note that our Dirac operator obeys
\begin{equation}
\tau_{1,2}\epsilon[D\!\!\!\!/(\tau_3\mu_I)+m]\epsilon\tau_{1,2}
                 =[D\!\!\!\!/(\tau_3\mu_I)+m]^\dagger,
\end{equation}
so $i\bar{\chi}\epsilon\tau_{1,2}\chi$ are Goldstone candidates. These are
linear combinations of $\pi^+$ and $\pi^-$ creation operators which means that
if spontaneous breaking of the remnant flavour symmetry should occur, one
linear combination of $\pi^\pm$ will become a Goldstone boson while the 
orthogonal linear combination will develop a vacuum expectation value -- a
charged pion condensate. We note, in passing, that in the limit of massless
quarks
\begin{equation}                                                        
\tau_{1,2}D\!\!\!\!/(\tau_3\mu_I)\tau_{1,2}=-D\!\!\!\!/(\tau_3\mu_I)^\dagger,
\end{equation}                                                 
and we have 2 additional Goldstone boson candidates, $\bar{\chi}\tau_{1,2}\chi$,
and if spontaneous symmetry breaking does occur we will have 2 Goldstone bosons
rather than 1.

Since, in order to observe spontaneous symmetry breaking on a finite lattice,
one needs to add a small explicit symmetry breaking term in the direction 
defined by the condensate, we choose to work with the fermion action
\begin{equation}
S_f=\sum_{sites} \left[\bar{\chi}[D\!\!\!\!/(\tau_3\mu_I)+m]\chi
                   + i\lambda\epsilon\bar{\chi}\tau_2\chi\right]
\end{equation}
where the term proportional to the (small) parameter $\lambda$ serves this
purpose. The Dirac operator now has the determinant
\begin{equation}
\det[D\!\!\!\!/(\tau_3\mu_I) + m + i\lambda\epsilon\tau_2]
               =\det[{\cal A}^\dagger{\cal A}+\lambda^2]
\end{equation}
where we have defined 
\begin{equation}
{\cal A} \equiv D\!\!\!\!/(\mu_I)+m.
\end{equation}
(Note that this is a $1 \times 1$ matrix in the flavour space on which the
$\tau$s act.) We see that adding this symmetry breaking term has the effect of
rendering the determinant strictly positive, which enables us to use the
hybrid molecular-dynamics (HMD) algorithm to simulate this theory. Note that
this theory has 8 continuum flavours. We use the HMD method to take the
required fourth root of the determinant reducing this to 2 continuum flavours.
For the purpose of simulation, it is convenient to multiply the Dirac operator
on the left by the matrix diag$(1,-\epsilon)$ and on the right by the matrix
diag$(1,\epsilon)$. The transformed matrix $\widetilde{\cal M}$, has the same
determinant as the original Dirac operator, and
$\widetilde{\cal M}^\dagger\widetilde{\cal M}$ is block diagonal, with the
upper and lower blocks having the same determinant. This means that we
use `noisy' fermions and generate Gaussian noise for both upper and lower
components of $\widetilde{\cal M}\dot{\chi}$, but only keep the upper
components of $\dot{\chi}$ after the inversion. Thus we still have only 8
flavours in the quadratic formulation. This is completely analogous to the
odd-even lattice separation which prevents further species doubling in
staggered lattice QCD at zero chemical potential.

Quantities we measure include the chiral condensate,
\begin{equation}
\langle\bar{\psi}\psi\rangle \Leftrightarrow \langle\bar{\chi}\chi\rangle,
\end{equation}
the charged pion condensate
\begin{equation}
i\langle\bar{\psi}\gamma_5\tau_2\psi\rangle  \Leftrightarrow 
i\langle\bar{\chi}\epsilon\tau_2\chi\rangle
\end{equation}
and the isospin density
\begin{equation}
j_0^3 = \frac{1}{V}\left\langle{\partial S_f \over \partial\mu_I}\right\rangle.
\end{equation}
Here we have included both the lattice and continuum versions of the
condensates. To get this simple continuum form for the charged pion condensate
requires absorbing a factor of $\xi_5$ (the flavour analogue of $\gamma_5$)
into the definition of the $d$-quark field.

\section{Lattice simulations at zero temperature}

We have simulated $N_f=2$ lattice QCD at finite $\mu_I$ on an $8^4$ lattice
at an intermediate coupling $\beta=6/g^2=5.2$. This $\beta$ was chosen since
it represents an approximate lower bound to estimates of the finite temperature
transition value for $N_t=4$ in the chiral limit. This was used to keep finite
volume effects at acceptable levels. We performed simulations at 2 different
quark masses ($m=0.025$ and $m=0.05$) to see that varying the mass did not
affect the qualitative behaviour of the theory and that we understood the
effects of changing the quark mass.

At $m=0.025$, we performed runs, each of 2000 molecular-dynamics time units
in length, at 17 different $\mu_I$ values ($0 \le \mu_I \le 2$) for each of 
$\lambda=0.0025$ and $\lambda=0.005$. Using 2 $\lambda$ values, both chosen to
be much less than $m$, enabled us to extrapolate to the $\lambda=0$ limit,
which is our ultimate interest. (We also ran at $\mu_I=3.0$, $\lambda=0.005$ to
check saturation.)

\begin{figure}[htb]
\epsfxsize=6in
\centerline{\epsffile{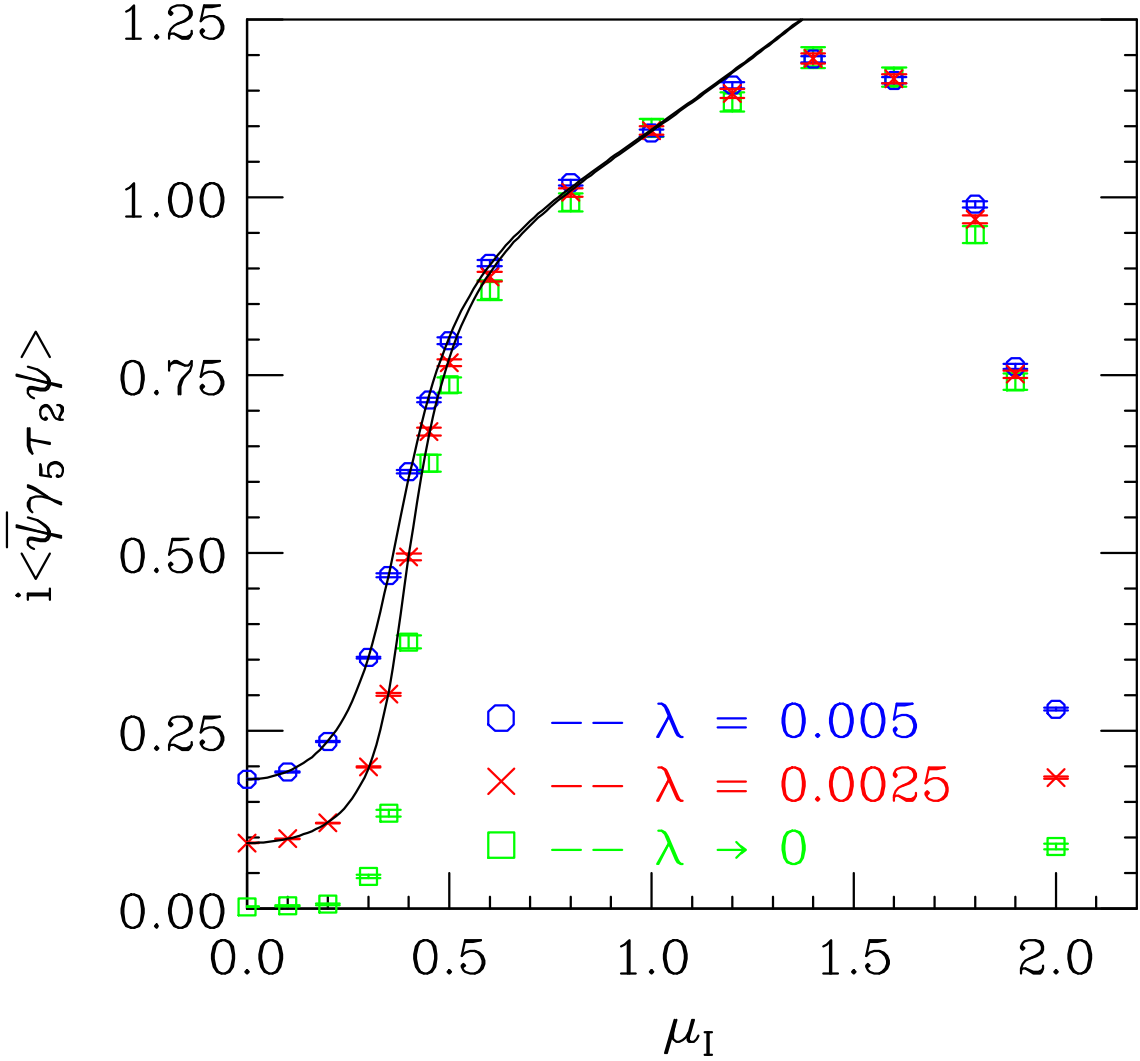}}
\caption{Charged pion condensate as a function of $\mu_I$ for $\lambda=0.0025$,
$\lambda=0.005$ and $\lambda \longrightarrow 0$. The curves are fits of the
finite $\lambda$ measurements to the scaling forms defined in the text.}
\label{fig:pi0.025}
\end{figure}
The charged pion condensates, $i\langle\bar{\psi}\gamma_5\tau_2\psi\rangle$
from these simulations are presented in figure~\ref{fig:pi0.025} as functions of
$\mu_I$, along with a linear extrapolation to $\lambda=0$. This extrapolation
strongly suggests that the $\lambda \longrightarrow 0$ condensate vanishes 
for $\mu_I < \mu_c \sim 0.3-0.45$, above which it is finite. This would indicate
a phase transition to a phase in which $I_3$ symmetry is broken spontaneously
by a charged pion condensate, with an associated Goldstone mode.

This behaviour is predicted by the effective (chiral) Lagrangian analyses of
Son and Stephanov, which also predict that the transition should be second
order with mean-field exponents \cite{sonstep}.  We fit our extrapolated 
`data' to the critical scaling form
\begin{equation}
i\langle\bar{\psi}\gamma_5\tau_2\psi\rangle = const\,(\mu_I-\mu_c)^{\beta_m}
\end{equation}
for $\mu_I > \mu_c$ close to the transition. Fitting to this form over the
range $0.4 \le \mu_I \le 1.0$, we find $\mu_c=0.394(1)$, $\beta_m=0.230(9)$
and $const=1.23(2)$ at a $62\%$ confidence level. This appears inconsistent
with mean field scaling for which $\beta_m=\frac{1}{2}$, and closer to the 
tricritical scaling for which $\beta_m=\frac{1}{4}$. Indeed, good fits to a
tricritical scaling form over this range of $\mu_I$ and both $\lambda$s
can be obtained (confidence level $25\%$).

However, as we have noted in our paper on the quenched theory \cite{quenched}
such fits can be deceptive and can be because the true scaling behaviour of
these theories is best described by the scaling forms given by effective
Lagrangian analyses. When this form of scaling is described in terms of
$\mu_I-\mu_c$ rather than the natural scaling variables these forms imply, the
true scaling window is very narrow. Outside this window these theories can 
appear to scale with $\beta_m$ which is half the true value when analyzed in
terms of $\mu_I-\mu_c$. We now introduce the scaling forms (equations of state)
suggested by such effective Lagrangians, both of which give mean-field scaling
behaviour.

The form for the equation of state suggested by the lowest order tree-level
analysis of effective Lagrangians of the non-linear sigma model type 
\cite{sonstep} is given in terms of $\alpha$ which minimizes the effective
potential
\begin{equation}
{\cal E} = - a \: \mu^2 \: \sin^2(\alpha) - b \: m \: \cos(\alpha)
           - b \: \lambda \: \sin(\alpha)
\label{eqn:nls1}
\end{equation}
in terms of which
\begin{equation}
i\langle\bar{\psi}\gamma_5\tau_2\psi\rangle = b \: \sin(\alpha)
\label{eqn:nls2}
\end{equation}
\begin{equation}
\langle\bar{\psi}\psi\rangle = b \: \cos(\alpha)
\label{eqn:nls3}
\end{equation}
and
\begin{equation}
j_0^3 = 4 \: a \: \mu \: \sin^2(\alpha).
\label{eqn:nls4}
\end{equation}
$b$ is given in terms of $\mu_c$ and $a$, namely
\begin{equation}
b = \frac{2}{m} \: a \: \mu_c^2
\end{equation}

The form for the equation of state which is derived from an effective
Lagrangian of the linear sigma model type is obtained by extracting the values
of $R$ and $\alpha$ which minimize the effective potential
\begin{equation}
{\cal E} = \frac{1}{4}\,R^4 - \frac{1}{2}\,a\,R^2 
         - \frac{1}{2}\,b\,\mu^2\,\sin^2(\alpha) 
         - c\,m\,R\,\cos(\alpha) - c\,\lambda\,R\,\sin(\alpha)
\label{eqn:ls1}
\end{equation}
in terms of which
\begin{equation}
i\langle\bar{\psi}\gamma_5\tau_2\psi\rangle = c \: R \: \sin(\alpha)
\label{eqn:ls2}
\end{equation}
\begin{equation}
\langle\bar{\psi}\psi\rangle = c \: R \: \cos(\alpha)
\label{eqn:ls3}
\end{equation}
and
\begin{equation}
j_0^3 =2 \: b \: \mu \: R^2 \sin^2(\alpha).
\label{eqn:ls4}
\end{equation}
$c$ is given in terms of $\mu_c$ by
\begin{equation}
c = { b \: \mu_c^2 \over m} \sqrt{a + b \: \mu_c^2}.
\end{equation}

Finally, the tricritical scaling form which we use for comparison, and which
does not yield mean-field behaviour, is expressed in terms of the value of
$\phi$ which minimizes the effective potential
\begin{equation}
{\cal E} = \frac{1}{6}\,\phi^6 - \frac{1}{3}\,c\,\lambda\,\phi^3 
         - \frac{1}{2}\,(\mu_I-\mu_c)\,\phi^2 - b\,\lambda\,\phi
\end{equation}
in terms of which,
\begin{equation}
i\langle\bar{\psi}\gamma_5\tau_2\psi\rangle = b\,\phi+\frac{1}{3}\,c\,\phi^3
\end{equation}
and
\begin{equation}
j_0 = a\,\phi^2.
\end{equation}

We fit our measurements of the charged pion condensate for $0 \le \mu_I \le 1$
and both $\lambda$s to the linear sigma model form of equation~\ref{eqn:ls2}. We
obtained a fit with $\mu_c=0.4036(5)$, $a=0.52(1)$, $m=0.0253(1)$ with 
$\chi^2/dof=2.2$. Although worse than the tricritical fit, we note that it
allows a reasonable fit for $\mu_I < \mu_c$ in addition to $\mu_I > \mu_c$
which the tricritical fit did not. $\mu_c=0.426(3)$ for the tricritical fit,
which is close to the value expected for the apparent tricritical scaling
produced by such sigma model scaling. One of the reasons the tricritical fit
gives better results is that it includes a second symmetry breaking interaction,
cubic in the order parameter, which rounds off the curve as $\mu_I$ approaches
$1.0$ allowing a better fit to the `data'. Such cubic terms could be 
incorporated equation~\ref{eqn:ls1}. However, now there is not one such term but
several, which is why this possibility was not considered. Finally, as we shall
see below, this fit makes good predictions for the chiral condensate and the
isospin density. 

\begin{figure}[htb]
\epsfxsize=6in
\centerline{\epsffile{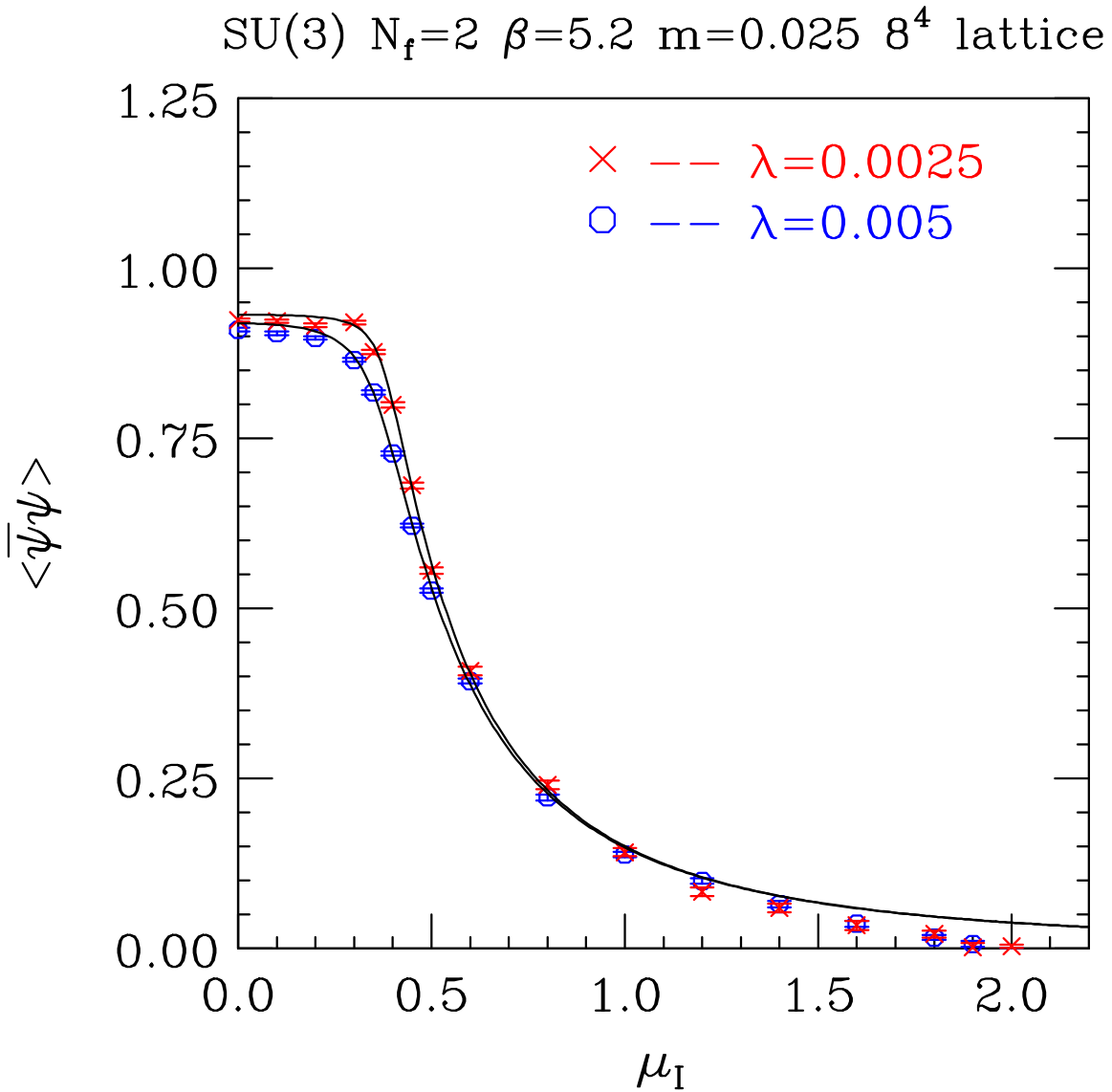}}
\caption{Chiral condensate as a function of $\mu_I$ for $\lambda=0.0025$ and
$\lambda=0.005$.}\label{fig:pbp0.025}
\end{figure}
In figure~\ref{fig:pbp0.025} we present the chiral condensate for the same set
of simulations. The general characteristics of these graphs are that
$\langle\bar{\psi}\psi\rangle$ remains roughly constant for $\mu_I < \mu_c$,
after which it drops rapidly, approaching zero at large $\mu_I$. Although
this can qualitatively be thought of as the condensate rotating from the
chiral to the isospin-breaking direction, it is not a simple rotation, since
the magnitude of the total condensate increases up until saturation effects 
take over. This contrasts with the predictions of lowest order effective
Lagrangians of the non-linear sigma model type where it is a simple rotation.
However, in the case of 2-colour QCD at finite quark-number chemical
potential, whose effective Lagrangian is structurally very similar to that for
the theory at hand, the effective Lagrangian/chiral perturbation theory
calculations have been extended to next-to-leading order \cite{stv}. Here,
although mean field scaling survives, the rotation of the condensate is
accompanied by a rescaling. Such behaviour is reproduced at tree level by
effective Lagrangians of the linear sigma model variety which we use for our
fits to the pion condensate. The predictions these fits make for the behaviour
of the chiral condensate (equation~\ref{eqn:ls3}), have been superimposed on
the measurements of figure~\ref{fig:pbp0.025}. Except for small departures at
small $\mu_I$, which we attribute to the fact that $m$ for the fit differs
slightly from the true $m=0.025$, these predictions are very good until the
effects of saturation start to dominate.

\begin{figure}[htb]
\epsfxsize=6in
\centerline{\epsffile{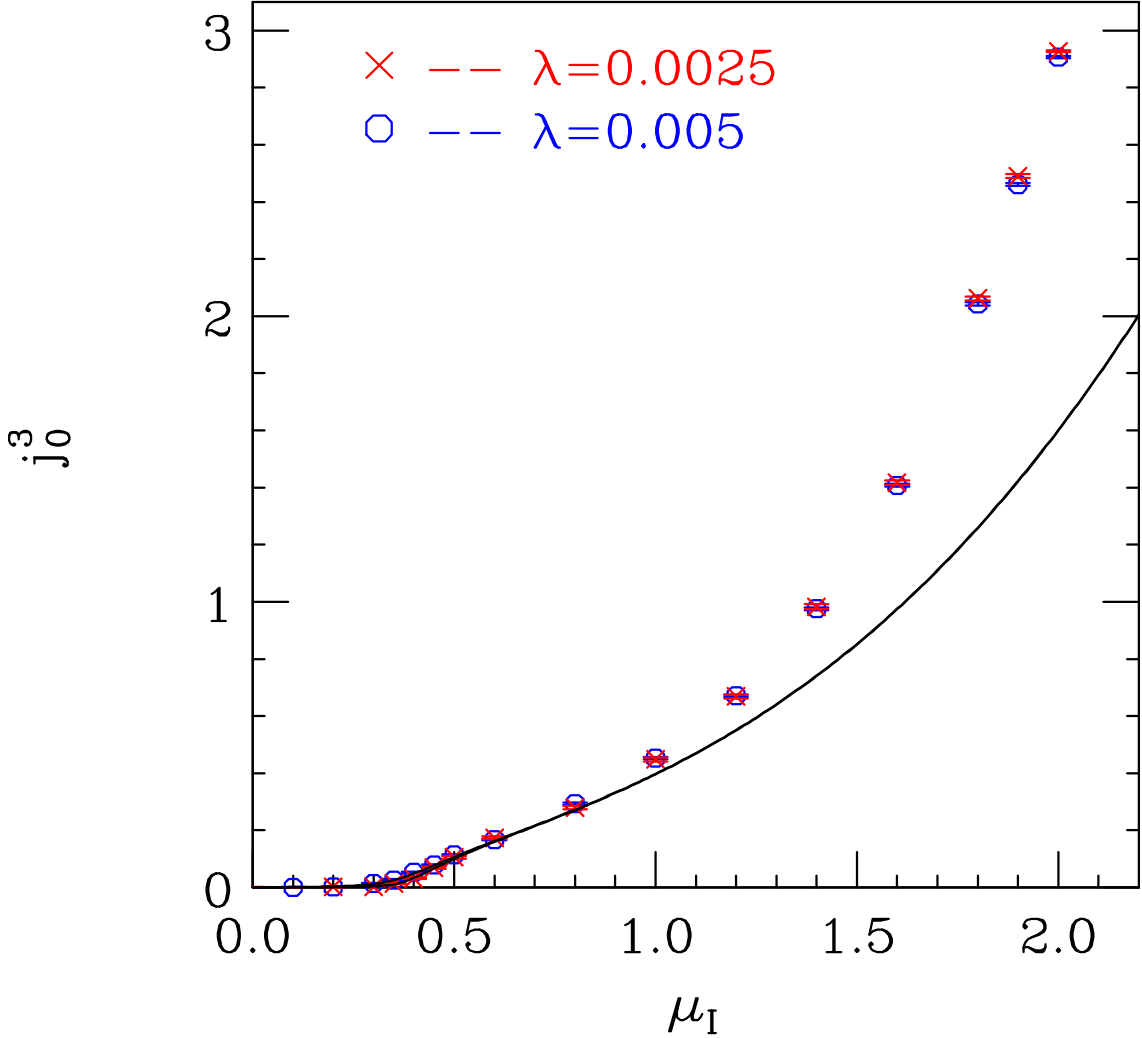}}
\caption{Isospin($I_3$) density as a function of $\mu_I$ for $\lambda=0.0025$ 
and $\lambda=0.005$.}\label{fig:j0.025}
\end{figure}
Figure~\ref{fig:j0.025} presents the isospin ($I_3$) density as a function of
$\mu_I$. While the 2 condensates are normalized to 4 flavours for comparison
with previous ($\mu_I=0$) simulations, $j_0^3$ is normalized to $8$ flavours,
which is the natural normalization for staggered quarks. Qualitatively, we note
that $j_0^3$ is close to zero for $\mu_I < \mu_c$, rises slowly (in comparison
with the pion condensate) up to $\mu_I \sim 1$, after which it starts to rise
more rapidly reaching its saturation value of $3$ (1/2 for each of 3 colours
and 2 `flavours' per site) due to fermi statistics, at $\mu_I \sim 2$. 
(In fact measurements made at $\mu_I = 3.0$ give a value consistent with $3$).
Saturation is clearly a finite lattice spacing effect and hence requires no
further discussion. We also note that there is very little $\lambda$ dependence.
The predictions of equation~\ref{eqn:ls4} using the parameters of our fit are
superimposed on our `data' and show good agreement out to $\mu_I \approx 0.8$.
Both `data' and fit are approximately linear in $\mu_I$ over this range.


As also noted by Son and Stephanov, measuring $j_0^3$ as a function of $\mu_I$
at $T=0$ and constant volume (since $\beta$ is constant) yields the pressure
$p$ and energy density $\epsilon$ as functions of $\mu_I$, since
\begin{eqnarray}
     p   &=& \int_{\mu_c}^{\mu_I} j_0^3 d \mu_I             \\
\epsilon &=& \int_0^{j_0^3} \mu_I d j_0^3 
\end{eqnarray}
for $\mu_I > \mu_c$ and zero for $\mu_I < \mu_c$.
In the region where $j_0^3 \approx const\,(\mu_I-\mu_c)$ these yield
\begin{eqnarray}                                                              
         p         &=& \frac{1}{2}\,const\,(\mu_I - \mu_c)^2                \\ 
     \epsilon      &=& \frac{1}{2}\,const\,(\mu_I^2 - \mu_c^2)              \\
{p \over \epsilon} &=& {\mu_I - \mu_c \over \mu_I + \mu_c}
\end{eqnarray}                                                                
The last of these equations is a form of the equation of state for this system
in the neighbourhood of $\mu_c$. Clearly we could extend each of these
expressions beyond the scaling region by interpolating the `data' of
figure~\ref{fig:j0.025} and performing the relevant integrals analytically or
numerically.

\begin{figure}[htb]
\epsfxsize=6in
\centerline{\epsffile{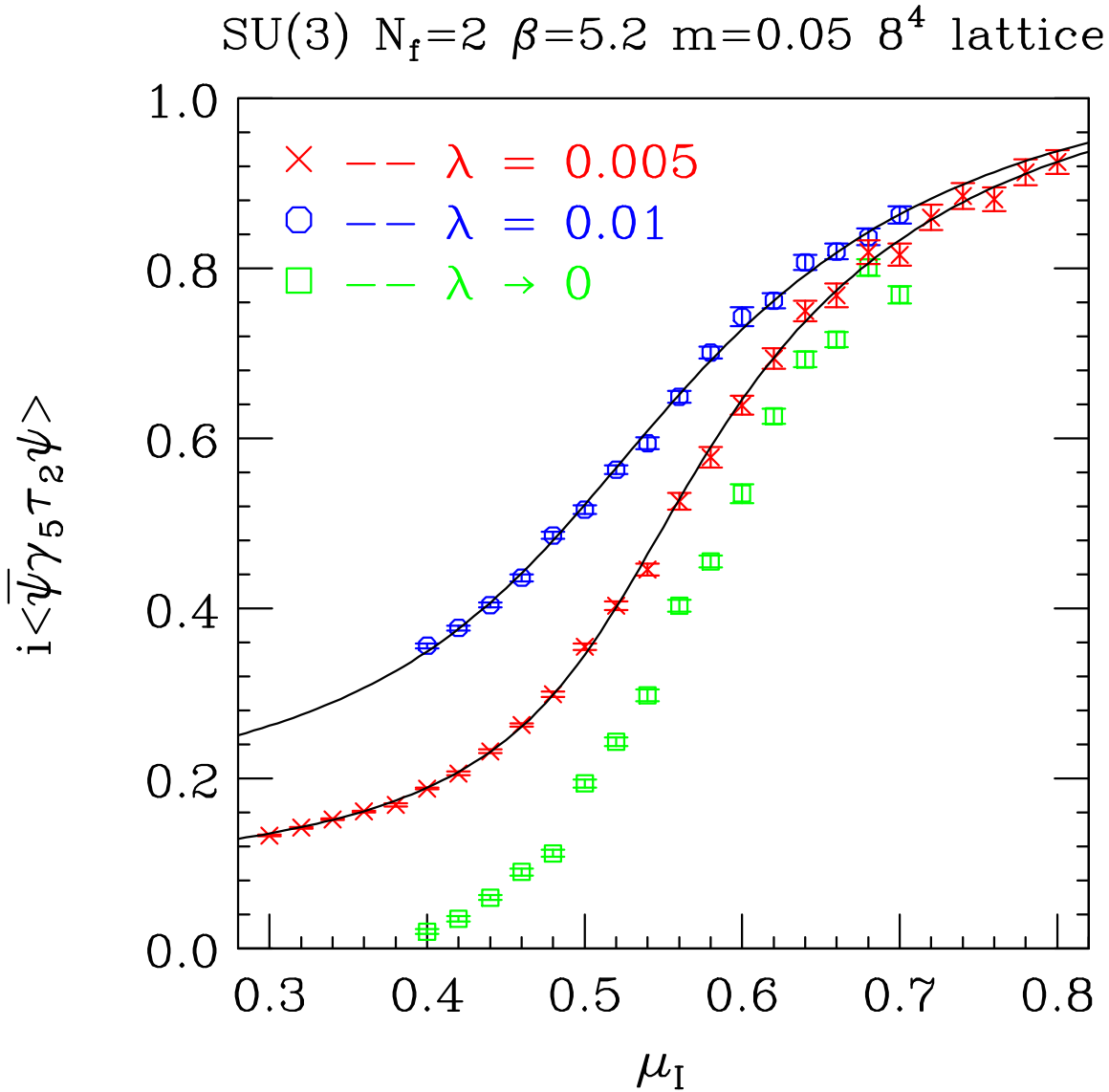}}
\caption{Pion condensates as functions of $\mu_I$ for $\lambda=0.005$,
$\lambda=0.01$ and a linear extrapolation to $\lambda=0.0$. The lines are the 
tricritical fit described in the text.}\label{fig:pi0.05}
\end{figure}
We performed similar simulations at the same coupling $\beta=5.2$ and mass
$m=0.05$ at $\lambda=0.005$ and $\lambda=0.01$. Here we concentrated on the
neighbourhood of the phase transition and used more closely spaced (in $\mu_I$)
points with somewhat lower statistics. For $\mu_I \gtrsim \mu_c$ we find
acceptable fits of the pion condensate to to both the non-linear sigma model
scaling form and to the tricritical scaling form ($34\%$ and $40\%$ 
respectively). The fit to non-linear sigma model effective Lagrangian scaling
(equation~\ref{eqn:nls2}) enables one to extend this to low $\mu_I$. A fit of
the `data' over the complete range over which we have measurements at both
$\lambda$ values --- $0.4 \le \mu_I \le 0.7$ --- yields $\mu_c=0.569(2)$,
$a=0.0868(9)$ and $m=0.0534(3)$, with $\chi^2/dof=1.5$ for the non-linear sigma
model form. Even though this fit is worse than those restricted to 
$\mu_I \gtrsim \mu_c$, we feel that the ability to fit $\mu_I < \mu_c$ in
addition to $\mu_I > \mu_c$ makes the argument for this form of fit more
compelling. Not only do we see qualitative consistency with the smaller mass
results, but our measured values of $\mu_c$ are consistent with the
expectation that $\mu_c(m=0.05)=\sqrt{2}\mu_c(m=0.025)$ which would be true
if, indeed, $\mu_c=m_\pi$, from PCAC. This `data' for the pion condensate with
the scaling fits superimposed is plotted in figure~\ref{fig:pi0.05}. We note
that the fits appear to have validity beyond the range of $\mu_I$ over which
the fits were performed.

\section{Simulations at finite temperature}

We have performed simulations of QCD at finite $\mu_I$ and finite temperature
on an $8^3 \times 4$ lattice, with $m=0.05$. Most of these simulations were
performed at $\lambda=0.005$ and $\lambda=0.01$, i.e. $\lambda << m$, with the
objective of obtaining information about the $\lambda=0$ limit. The goal of
these simulations is to map out the region of the $(\beta, \mu_I)$ and hence
the $(T, \mu_I)$ plane where $I_3$ is spontaneously broken by a charged pion
condensate, and determine the nature of the phase transitions which demarcate
its boundaries.

The first of these simulations was performed at a fixed, large (but well below
saturation) value of $\mu_I$. The value chosen was $\mu_I=0.8$. At $\beta$ low
enough to approximate zero temperature on an $N_t=4$ lattice, the system is
in the phase where $I_3$ is spontaneously broken by a (large) pion condensate.
As $\beta$ is increased we eventually reach a value $\beta=\beta_c$ at which
this condensate evaporates. For $\beta > \beta_c$ we are in the phase where
the pion condensate vanishes for $\lambda \longrightarrow 0$ and $I_3$ is
unbroken. A single $\lambda$ value, $\lambda=0.005$ was used for these runs.

\begin{figure}[htb]
\epsfxsize=6in                                                                  
\centerline{\epsffile{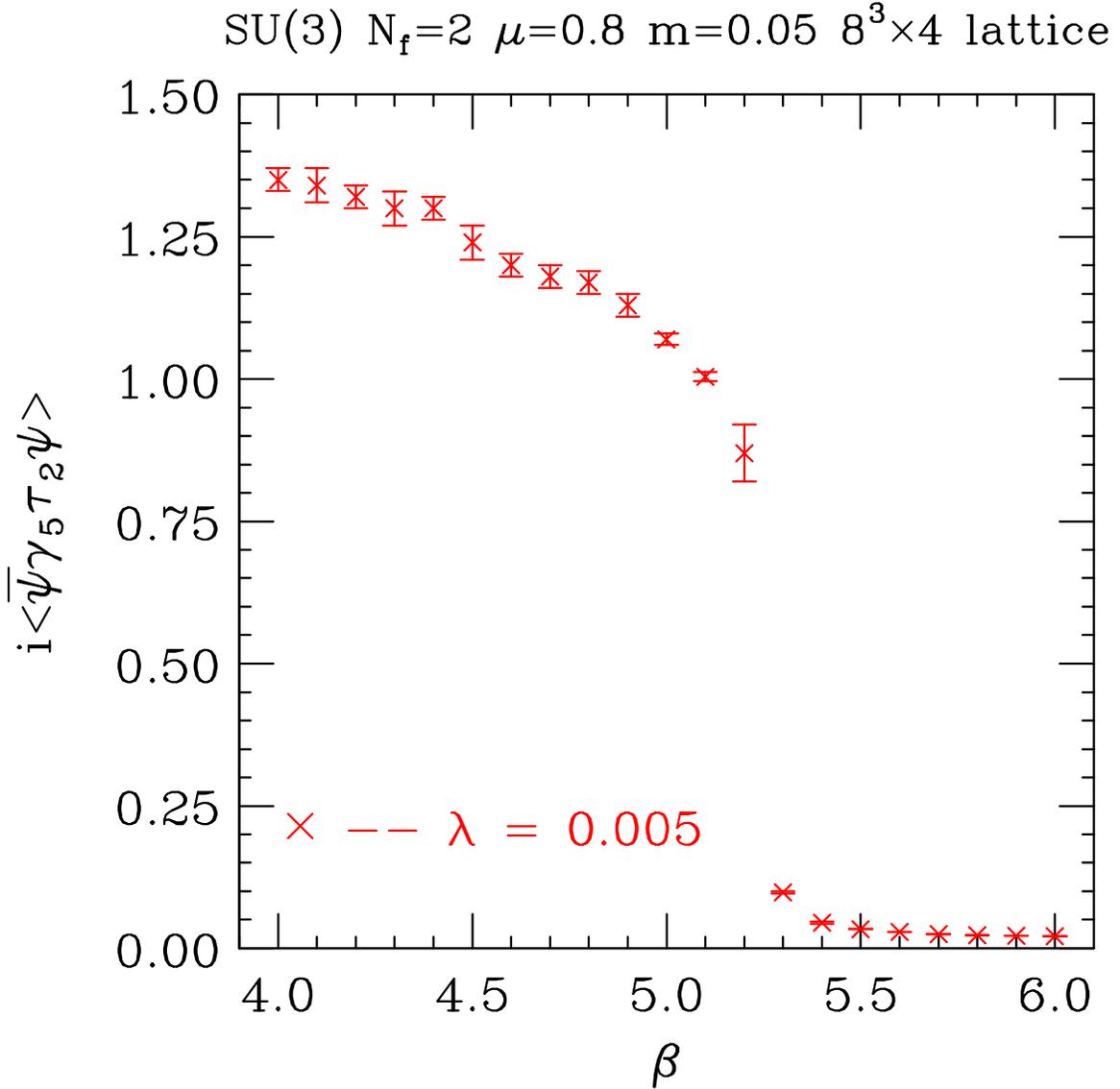}}                                     
\caption{Pion condensate as a function of $\beta$ for $m=0.05$, $\lambda=0.005$,
and $\mu_I=0.8$ on an $8^3 \times 4$ lattice.}\label{fig:pi_mu8}
\end{figure}
Figure~\ref{fig:pi_mu8} shows the $\beta$ dependence of the pion condensate
for these simulations. We see that for $\beta \le 5.2$ the condensate is large.
Between $\beta=5.2$ and $\beta=5.3$ the condensate drops by an order of 
magnitude, and is small enough for $\beta \ge 5.3$ that we are safe to assume 
that it would vanish in the $\lambda \longrightarrow 0$ limit. This drop is
so precipitous that we suspect that it is first order, although we really need
a larger lattice to confirm this.

\begin{figure}[htb]
\epsfxsize=6in
\centerline{\epsffile{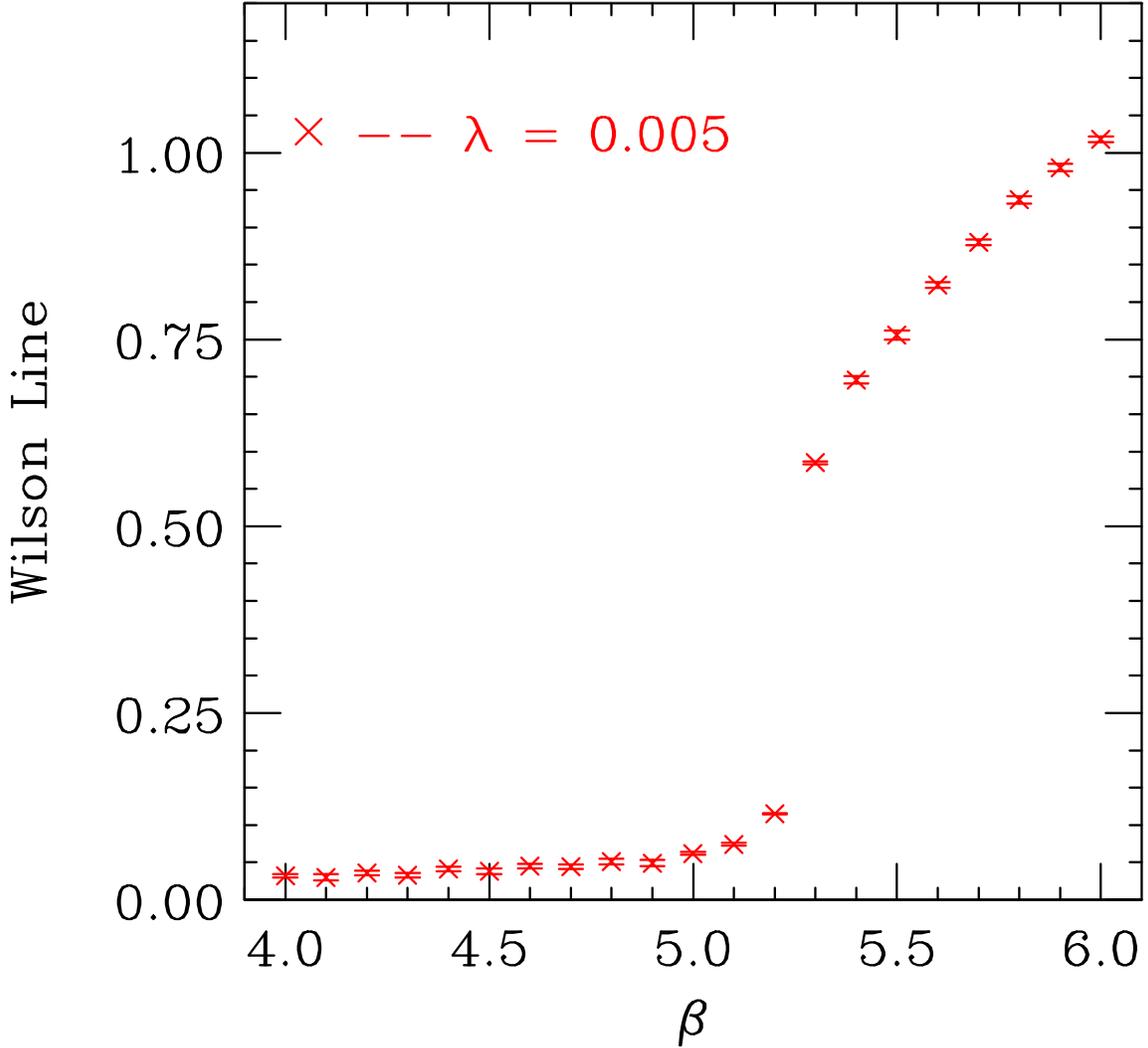}}
\caption{Wilson linear a function of $\beta$ for $m=0.05$, $\lambda=0.005$,
and $\mu_I=0.8$ on an $8^3 \times 4$ lattice.}\label{fig:wl_mu8}                
\end{figure}
We have also measured the Thermal Wilson Line (Polyakov Loop) during these 
runs. These measurements are shown in figure~\ref{fig:wl_mu8}. For 
$\beta \le 5.2$, the Wilson line is small indicating confinement. For
$\beta \ge 5.3$, the Wilson line becomes large indicating deconfinement. The
jump between $\beta=5.2$ and $\beta=5.3$ is again great enough to suggest a
first order transition. This behaviour of the Wilson Line indicates that this
is the temperature-driven deconfinement transition.

\begin{figure}[htb]
\epsfxsize=6in
\centerline{\epsffile{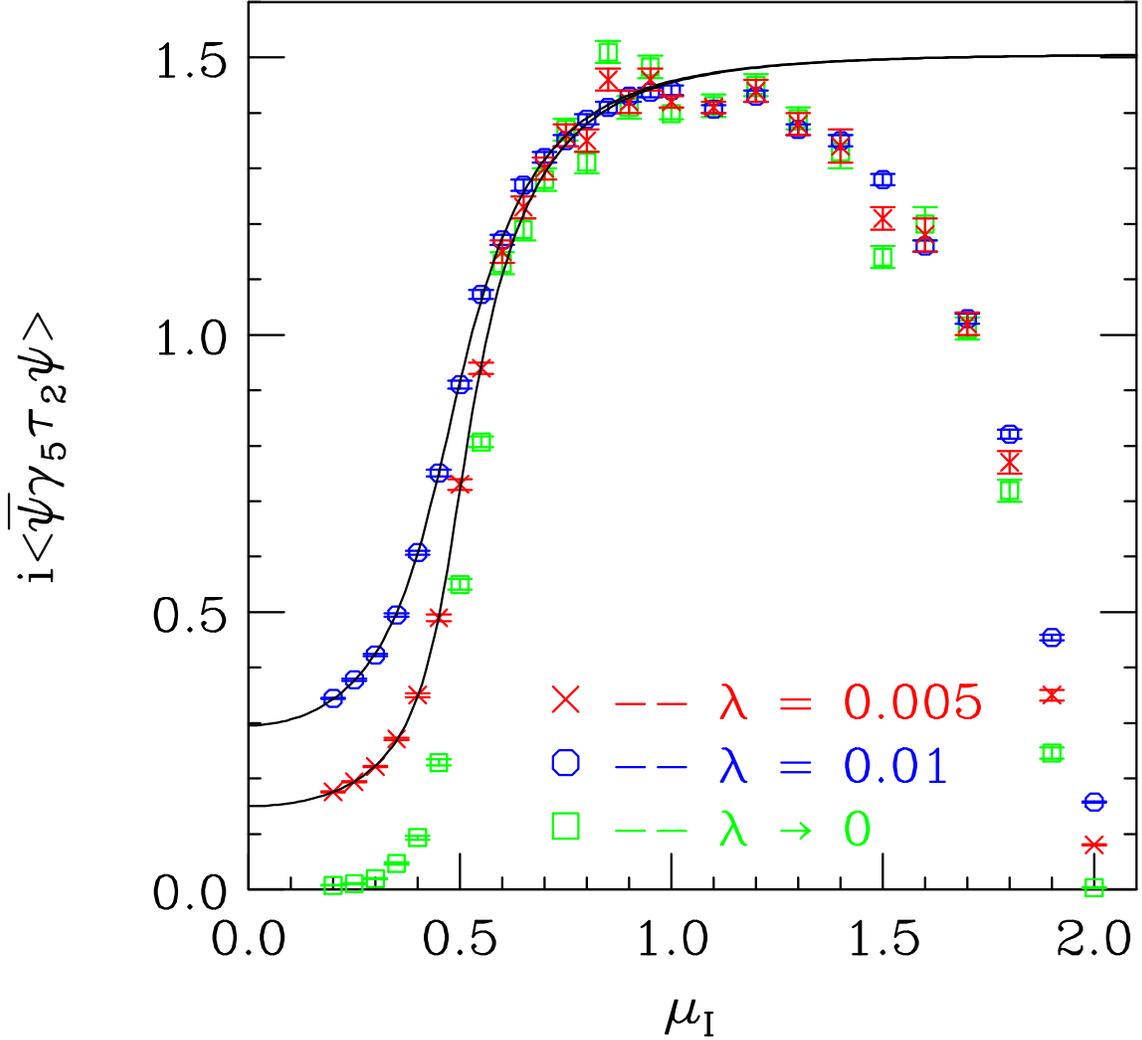}}
\caption{Pion condensates as functions of $\mu_I$ for $\lambda=0.005$,
$\lambda=0.01$ and a linear extrapolation to $\lambda=0.0$ on an $8^3 \times 4$
lattice at $\beta=4.0$. The lines are the tricritical fit described in the 
text.} \label{fig:pi0.05b4_4}
\end{figure}
We have also run simulations on an $8^3 \times 4$ lattice with $\beta=4.0$
which gives us the low temperature behaviour. We chose $m=0.05$ again and ran
at $\lambda=0.005$ and $\lambda=0.01$ for $0.20 \le \mu_I \le 2.0$ which covers
both the transition from the $I_3$ symmetric phase to the phase where $I_3$ is
spontaneously broken, and the approach to saturation. Our measurements of the
pion condensate are shown in figure~\ref{fig:pi0.05b4_4}. Not surprisingly this
graph resembles those we obtained on an $8^4$ lattice since at $\beta=4.0$,
this lattice is essentially at zero temperature.

We fit the scaling behaviour of these measurements to the non-linear sigma
model scaling form of equation~\ref{eqn:nls2} for both $\lambda$ values and
$0.2 \le \mu_I \le 0.8$. The fit yielded $\mu_c=0.519(1)$, $a=0.1400(4)$ with
the quark mass fixed at $m=0.05$ at a confidence level of $41\%$, which is
very good. These fits are shown in figure~\ref{fig:pi0.05b4_4}. Again an
acceptable tricritical fit was possible (confidence level $48\%$), but 
only for $0.55 \le \mu_I \le 0.8$, yielding the expected larger estimate for 
$\mu_c$.

\begin{figure}[htb]
\epsfxsize=6in
\centerline{\epsffile{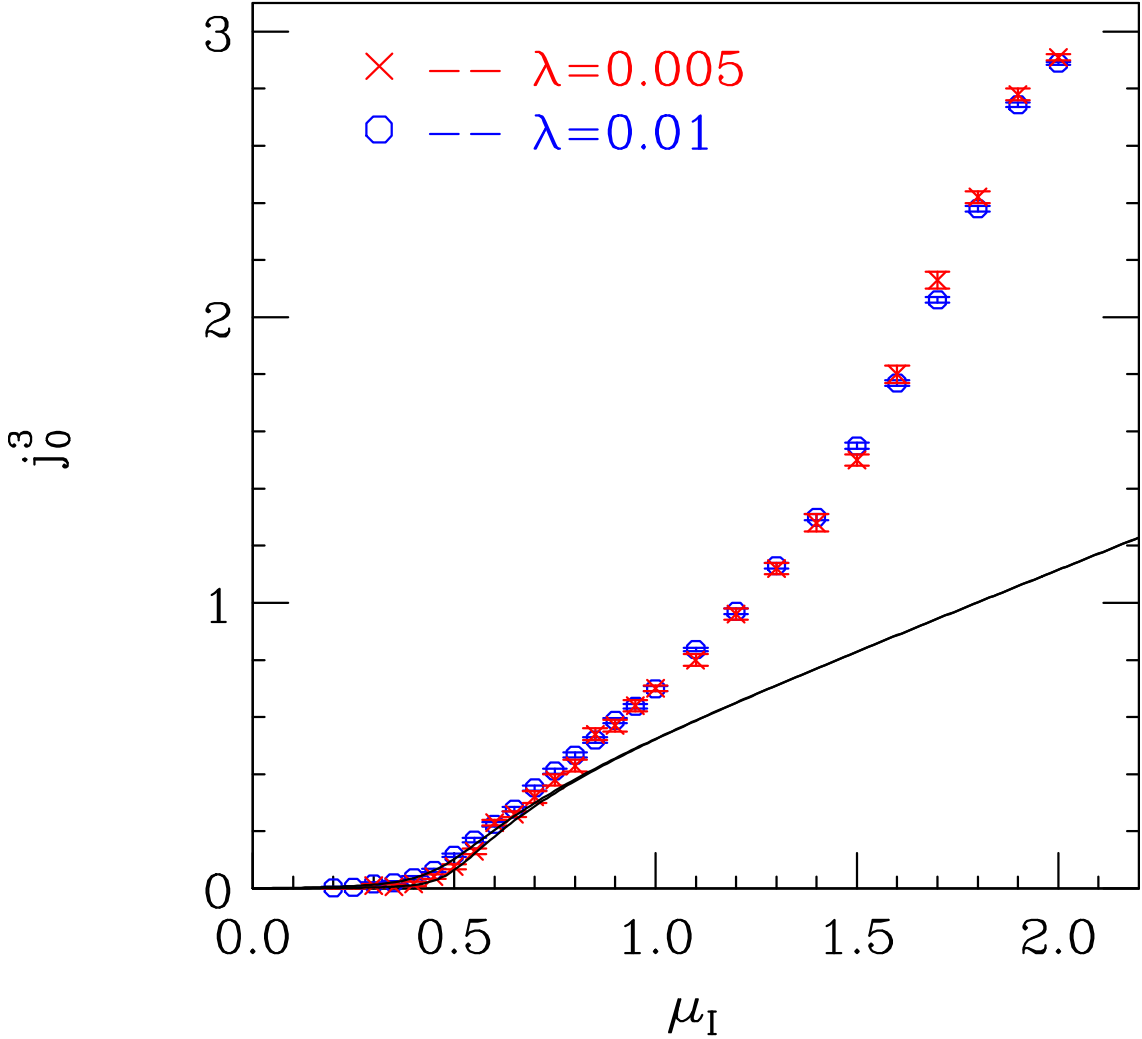}}
\caption{Isospin($I_3$) density as a function of $\mu_I$ for $m=0.05$,
$\lambda=0.005$ and $\lambda=0.01$ on an $8^3 \times 4$ lattice at $\beta=4.0$.}
\label{fig:j0.05b4_4}
\end{figure}
In figure~\ref{fig:j0.05b4_4} we present the isospin density from these runs. 
$j_0^3$ rises from zero at $\mu_I \sim \mu_c$. Again there is little $\lambda$
dependence. We have superimposed the form predicted from the fit to pion
condensate using equation~\ref{eqn:nls4} on these plots. These curves are in
reasonable agreement with the measured values up to $\mu_I \approx 0.7$. This
indicates that the scaling window for $j_0^3$ is slightly less than that for
the pion condensate. However, it is clear that the linear or near-linear
increase of this quantity with $\mu_I$ continues beyond the point where the
`data' and curves diverge. This is born out by the fitting $j_0^3$ to the form
\begin{equation}                         
j_0^3 = const\,(\mu_I - \mu_c)^{\beta_I}.
\label{eqn:jscale}                       
\end{equation}                                                                  
Fitting the $\lambda=0.005$ `data' over the range $0.5 \le \mu_I \le 1.1$
gives $\mu_c=0.467(13)$, $\beta_I=0.94(6)$ and $const=1.22(3)$ at a confidence
level of $18\%$ while fitting the $\lambda=0.01$ `data' yields
$\mu_c=0.382(9)$, $\beta_I=1.09(3)$, $const=1.20(1)$ at a confidence level of
$85\%$. These results are in excellent agreement with the effective Lagrangian
prediction $\beta_I=1$. This graph also indicates a crossover to a more rapid
increase at $\mu_I \sim 1.5$.

\begin{figure}[htb]
\epsfxsize=6in
\centerline{\epsffile{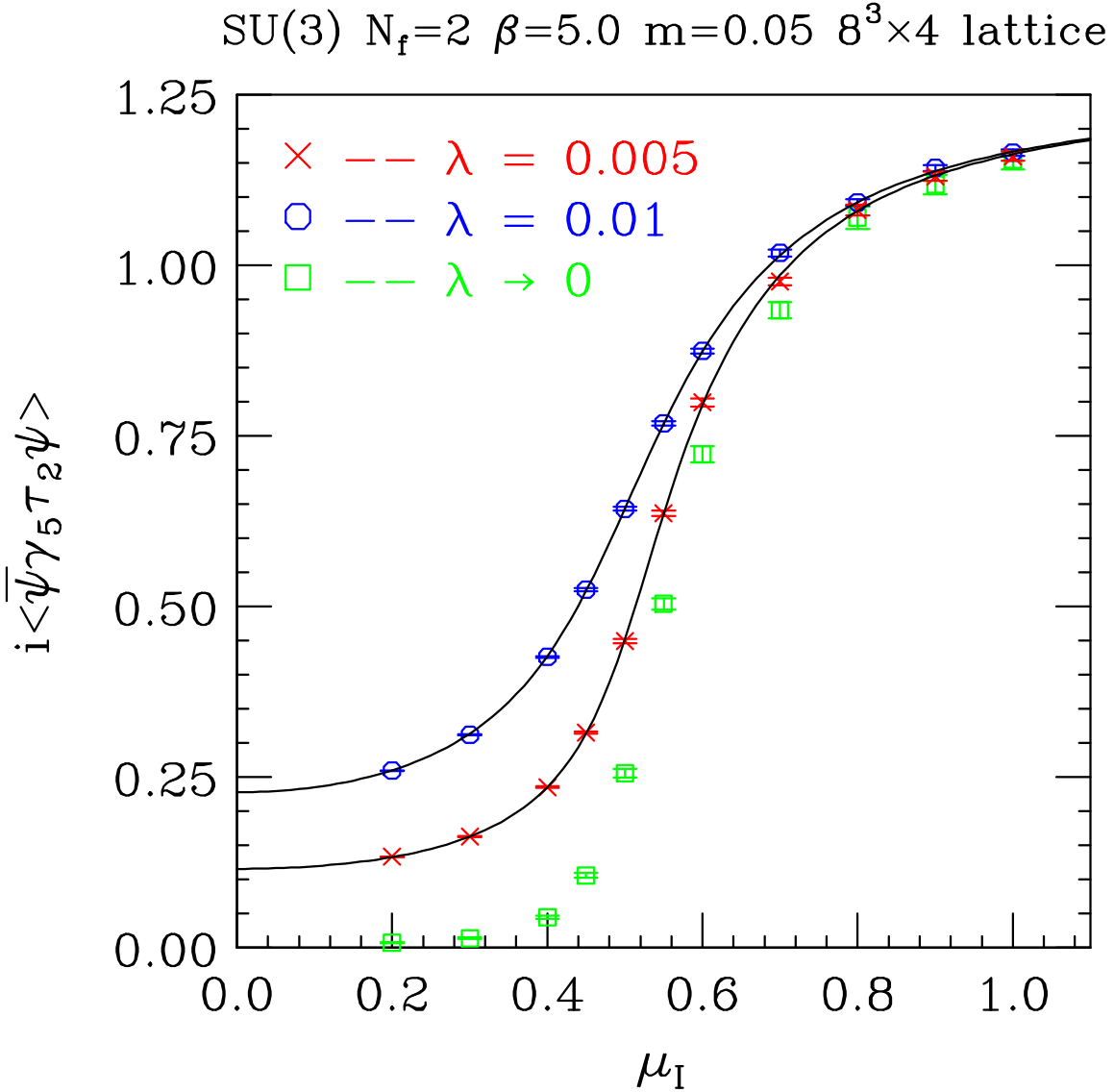}}
\caption{Pion condensates as functions of $\mu_I$ for $\lambda=0.005$,
$\lambda=0.01$ and a linear extrapolation to $\lambda=0.0$ on an $8^3 \times 4$
lattice, at $\beta=5.0$. The lines are the tricritical fit described in the
text.}\label{fig:pi0.05b5_4}
\end{figure}
Finally we have performed $m=0.05$ simulations on an $8^3 \times 4$ lattice at
$\beta=5.0$, which lies below $\beta_c$ at high $\mu_I$, while being large
enough that the effects of finite temperature should be apparent. Here again we
ran with $\lambda=0.005,0.01$. Here we used higher statistics (2000 time-units
per `run') but at fewer $\mu_I$ values. Our measurements of the charged pion
condensate are presented in figure~\ref{fig:pi0.05b5_4}. The transition from
low to high values of this condensate shows no sign of a first order transition.
The linear extrapolation to $\lambda=0$ gives values close to zero at low 
$\mu_I$, rising rapidly from zero above some $\mu_c \sim 0.5$. Here we find 
good fits of the `data' for both $\lambda$ values and $0.2 \le \mu_I \le 1.0$
to the non-linear sigma model form with $\mu_c=0.5521(5)$, $a=0.1040(4)$,
$m=0.0523(1)$ at a confidence level of $31\%$ and to the linear sigma model
form with $\mu_c=0.5513(5)$, $a=2.46(1)$, $b=0.0823(4)$, $m=0.0518(1)$ at
a confidence level of $48\%$. This second fit in included in 
figure~\ref{fig:pi0.05b5_4}. A tricritical fit is possible for $0.55 \le \mu_I
\le 0.8$ and has a confidence level of $28\%$.

Thus the line of phase transitions which bound the region in the $(\mu_I,T)$
plane within which isospin ($I_3$) is broken spontaneously by a charged pion
condensate, is second order for low temperatures and becomes first order at
high $\mu_I$. The second order segment of this line appears to have mean-field
critical exponents.

\section{Discussion and conclusions}

We have simulated QCD with 2 quark flavours ($u$,$d$) at a finite chemical
potential ($\mu_I$) for isospin, $I_3$. At zero temperature and intermediate
coupling ($\beta=6/g^2=5.2$) we found strong evidence for a second order
transition to a phase in which $I_3$ is spontaneously broken by a charged
pion condensate which also breaks parity, at $\mu_I=\mu_c$. 
The observed behaviour
is what is predicted by effective Lagrangian methods for $\mu_I$ appreciably
less than the value at which saturation, a lattice artifact, takes over. These
effective Lagrangian analyses predict that $\mu_c=m_\pi$. Since we have not
measured $m_\pi$ on these small lattices, all we have checked is that $\mu_c
\propto \sqrt{m}$ for the 2 quark masses which we use ($m=0.025$,$m=0.05$), and
found that this is true within the uncertainties of our measurements.
The critical scaling appears to be well described by an equation of state
suggested by these effective Lagrangian analyses, which means that the critical
point has mean-field critical exponents. However, tricritical behaviour cannot
be completely excluded. Such behaviour was discussed in detail in our paper
on the quenched theory \cite{quenched}.

We also measured the isospin density ($j_0^3$) and found that it increases from
zero, at or near $\mu_c$. The scaling behaviour appears to be linear close to
$\mu_c$, as is predicted by effective Lagrangian analyses \cite{sonstep} and
well described by the predictions given by the fits to the pion condensate,
within the scaling window. For larger $\mu_I$ values it starts to increase
considerably faster than linear. This is in qualitative agreement with the
expectation that $I_3$ density should increase as $\mu_I^3$ at large $\mu_I$.
We have not been able to determine if our observations are consistent with
this $\mu_I^3$ increase because of the effects of saturation which cause the
isospin density to approach $3$ at high $\mu_I$. Here we have indicated how
the measurement of $j_0^3$ enables one to obtain the pressure ($p$) and the
energy density ($\epsilon$) as functions of $\mu_I$, and given explicit
expressions for these quantities in the scaling regime.

The chiral condensate remains approximately constant for $\mu_I < \mu_c$.
Above $\mu_c$ it starts to fall approaching zero for large $\mu$. Again this
is in agreement with expectations, and the predictions from the fits to the
pion condensate. However, the expectation from lowest-order
effective Lagrangian tree-level analyses, that the chiral condensate simply
rotates into the direction of the charged pion condensate, is not true.
However, in closely related 2-colour QCD at finite quark-number chemical
potential, chiral perturbation theory calculations through next-to-leading
order show that, while scaling remains mean field, the condensate does not
merely rotate, but also rescales \cite{stv}. Since the structure of chiral
perturbation theory (effective Lagrangians) for the two theories is so
similar, we expect a similar result for QCD at finite $\mu_I$.

We have performed simulations at finite temperature ($T$) in addition to finite
$\mu_I$. In particular we have heated the system at fixed 
$\mu_I=0.8 > \mu_c(T=0)$ (in lattice units) by increasing $\beta$. On our
$N_t=4$ lattice there is some $\beta=\beta_c$ ($5.2 \le \beta_c \le 5.3$)
at which the charged pion condensate evaporates. This transition appears to
be first order. Such first order behaviour was predicted for $\mu_I$ large
enough by Son and Stephanov \cite{sonstep} who argued that at high $\mu_I$ the
fermions would effectively decouple, and the phase transition would be that
for pure glue, which is known to be first order. We note from our observations
that the situation is not quite this simple. The pure glue transition on an
$N_t=4$ lattice occurs at $\beta \approx 5.7$ \cite{columbia,ape}. Since our
observations place $\beta_c$ somewhat lower than this, and close to the value
of the $\mu_I=0$ finite temperature transition, the quarks {\it are} having an
effect. We also note that Son and Stephanov suggest that the first order
deconfinement transition at high temperature is distinct from the $I_3$
symmetry restoring transition. Our evidence is that these 2 transitions are
coincident. This means that the second order segment of the phase boundary can
be considered as driven purely by the chemical potential, which makes their
argument for $O(2)$ universality less compelling. We note that, in our paper on
2-colour QCD at finite $\mu$ and $T$ \cite{qcd2t}, we present an alternative
argument for the first-order finite-temperature transition which that theory
exhibits for large $\mu$.

In addition, we have simulated our $N_t=4$ system at fixed $\beta$, varying
$\mu_I$. In particular we have performed simulations at $\beta=4.0$ which is
at near-zero temperature, and $\beta=5.0$ where the system is clearly at a
finite temperature. Both of these simulations showed a second order transition.
Again the scaling was well described by the scaling forms suggested by effective
Lagrangian analyses which indicates that they have mean-field critical
exponents.

We have noted throughout this paper the similarity between 3-colour QCD at
finite $\mu_I$ and 2-colour QCD at finite quark-number chemical potential,
$\mu$. The correspondence is seen by identifying $\frac{1}{2}\mu_I$ with
$\mu$, the charged pion condensate with the diquark condensate and the isospin
density with the quark number density. This similarity is seen both in
simulations and in effective Lagrangian analyses. The expected position of the
zero temperature transition is the same $\mu_I = m_\pi$ 
($\mu =\frac{1}{2}m_\pi$). Its nature --- second order with mean-field
exponents --- is the same as predicted \cite{stv} and observed
\cite{qcd2it,newqcd2} for 2-colour QCD. In both systems the spontaneous
symmetry breaking is in the Goldstone mode (superfluid). At finite temperature
the condensate evaporates at a transition which is first order for $\mu_I$
($\mu$) large enough. This line of first order transitions softens to second
order and the line of second order transitions connects to the zero temperature
transition. Thus, we can use 2-colour QCD results at finite $\mu$ as a guide as
to QCD at finite $\mu_I$.

We are extending these simulations to a larger lattice ($12^3 \times 24$) and
weaker coupling where we hope to observe the expected mean-field transition
more clearly distinguished from the tricritical alternative 
and rule out the $O(2)$ alternative. This lattice will also enable us to
measure the spectrum of Goldstone and pseudo-Goldstone excitations as
functions of $\mu_I$. More extensive spectrum analyses at $\mu_I=0$ will give
us a more definitive scale for these phenomena, in addition to a value for
$m_\pi$ with which to compare $\mu_c$. Configurations will be stored so that
we can make other spectroscopy measurements at finite $\mu_I$. We will also
extend the finite temperature simulations to a $12^3 \times 6$ lattice since
it is difficult to determine the order of the continuum transitions from $8^3
\times 4$ lattices. In addition we will study the instantons at large $\mu_I$
since it is believed that instantons and their interactions have a relatively
simple structure at large isospin density, analogous to what has been
predicted for large quark-number density \cite{ssz}.

We are also starting lattice QCD simulations including both a chemical 
potential $\mu_I$ for isospin and a chemical potential $\mu_s$ for strangeness.
Here effective Lagrangian analyses have indicated that there is a competition
between the formation of pion and kaon condensates as $\mu_I$ and $\mu_s$ are
varied independently and that the boundary between the region with a pion 
condensate and that with a kaon condensate is a line of first order transitions
\cite{strange}.

Although our simulations have been limited to zero baryon number density, it
is interesting to know how much of this analysis is relevant to systems with
finite baryon number density and isospin density. If it does have relevance,
charged pion condensates could contribute to the equation-of-state of nuclear
matter and thus be important in understanding the physics of neutron stars
and perhaps large nuclei.

\section*{Acknowledgements}                    

DKS was supported by DOE contract W-31-109-ENG-38. JBK was supported in part
by an NSF grant NSF PHY-0102409. JBK wishes to thank D.~Toublan for many
useful discussions. DKS would like to thank R.~Pisarski for emphasizing the
importance of extracting pressure and energy density.

\end{document}